\documentclass[superscriptaddress,showpacs,amssymb,10pt,reprint,aps,prd,longbibliography,nofootinbib,floatfix]{revtex4-1}
\usepackage{graphicx,epsfig,amssymb} 
\usepackage{amsmath,amsfonts, times}
\usepackage{bm} 
\usepackage[normalem]{ulem}
\usepackage{epstopdf}
\usepackage[linktocpage,colorlinks]{hyperref}
\usepackage[caption=false]{subfig}
\usepackage[usenames]{color} 
\usepackage{mathrsfs} 
\usepackage{natbib}
\usepackage{soul}
\usepackage{subfig}
\usepackage[utf8x]{inputenc}
\usepackage{tikz}
\usetikzlibrary{decorations.pathmorphing}
\definecolor{coolblack}{rgb}{0.0, 0.18, 0.39}
\definecolor{darkred}{rgb}{0.5,0,0}
\definecolor{darkgreen}{rgb}{0,0.5,0}
\definecolor{darkblue}{rgb}{0,0,0.5}
\definecolor{lapislazuli}{rgb}{0.15, 0.38, 0.61}
\definecolor{venetianred}{rgb}{0.78, 0.03, 0.08}
\definecolor{bleudefrance}{rgb}{0.19, 0.55, 0.91}
\definecolor{dogwoodrose}{rgb}{0.84, 0.09, 0.41}
\hypersetup{colorlinks=true, citecolor=darkgreen, linkcolor=darkblue, 
urlcolor = blue}

\begin{document}
	\title{\large Black holes in higher-derivative Weyl conformal  gravity}

\author{Leandro A. Lessa}
	\email{leandrophys@gmail.com}
	\affiliation{Programa de Pós-graduação em Física, Universidade Federal do Maranhão, Campus Universitário do Bacanga, São Luís (MA), 65080-805, Brazil.}
	\author{Caio F. B. Macedo}
    \email{caiomacedo@ufpa.br}
	\affiliation{Programa de Pós-Graduação em Físic, Universidade Federal do Pará, 66075-110, Belém, PA, Brazil}
\author{Manoel M. Ferreira Jr.}
	\email{manoel.messias@ufma.br}
	\affiliation{Programa de Pós-graduação em Física, Universidade Federal do Maranhão, Campus Universitário do Bacanga, São Luís (MA), 65080-805, Brazil.}
	\affiliation{Departamento de Física, Universidade Federal do Maranhão, Campus Universitário do Bacanga, São Luís, Maranhão 65080-805, Brazil}
\begin{abstract}
We obtain exact black hole solutions for static and spherically symmetric sources in a Weyl conformal gauge theory of gravity. We consider a quadratic gravitational action built from the Weyl tensor within a dilation geometry. In a post-Riemannian formulation, we derive a Weyl conformal action for a scalar-vector-tensor theory, where the scalar degree of freedom originates from the high-curvature terms and the vectorial one stems from the Weyl non-metricity condition. Adopting a static, spherically symmetric geometry, the vacuum field equations for the gravitational, scalar, and Weyl fields are obtained. Under these conditions, we find a Mannheim-Kazanas-type black hole solution, whose Rindler acceleration term depends on the Weyl gauge coupling constant. Furthermore, we show that the original theory can recover the Einstein-Hilbert action with a positive cosmological constant plus a higher-derivative term with Horndeski-like terms through a spontaneous symmetry breaking triggered by the vacuum expectation value of the scalar field. The new solution in the theory without conformal symmetry presents new terms introduced by the residual Weyl symmetry corrections. We demonstrate that in a regime where the Planck mass suppresses the higher-derivative term, the Rindler term persists in the low-energy limit.

\end{abstract}
\date{\today}
\maketitle
\section{Introduction}
The foundational theories of modern physics are general relativity (GR), which describes gravitational interaction, and the standard model (SM), which governs electroweak and strong interactions. Both frameworks have proven remarkably successful in their respective domains: GR provides a geometric description of gravity, while the SM describes quantum interactions. GR is based on diffeomorphism and local Lorentz symmetries, whereas the SM relies on gauge symmetries. Despite decades of intense efforts to unify these theories within a single framework, all attempts have thus far failed - although some approaches, such as string theory, have achieved partial success.

One proposed path toward unification involves reformulating GR as a gauge theory. In 1918, Hermann Weyl first introduced a gauge-invariant version of gravity through a conformal quadratic vector-scalar theory, aiming to unify gravity with electromagnetism~\cite{weyl1918sitzungsber}. However, the theory was considered unviable and subsequently fell into obscurity \footnote{The initial criticism made by Einstein states that the spacing between atomic spectral lines changes under parallel transport, thus leading to a dependence of this distance on the trajectory history of each atom due to the non-metricity of Weyl geometry, and this contradicts the "second clock" experiment. However, as was later pointed out~\cite{Condeescu:2024nbs,ghilencea2025quantum}, this criticism is wrong. If the Weyl gauge covariance is respected, the conformal geometry is, in fact, metric, and parallel transport preserves the norm of vectors and clock rates}. Recent years have seen renewed interest in this approach, particularly due to work demonstrating how to circumvent Einstein's principal objections to Weyl gravity \cite{hobson2020weyl} (see Ref.~\cite{scholz2018unexpected} for a modern review).

    GR represents a geometric theory of gravitation based on a unique symmetric connection compatible with the metric - the Levi-Civita connection - where spacetime is described by a (pseudo-)Riemannian geometry characterized by its curvature tensor. Although extensively validated experimentally~\cite{will2006living,berti2015testing}, GR cannot be considered a complete theory of gravity as it faces significant challenges, including explaining the universe's dark sector, resolving physical singularities, and lacking a consistent quantum description. These limitations have motivated various modifications to the theory, among which Metric-Affine Gravity (MAG) has recently emerged as particularly promising~\cite{hehl1995metric,olmo2022conformal}. MAG extends the traditional formalism by treating both the metric and connection as independent dynamical fields. In this more general framework, the connection may contain antisymmetric components, introducing new geometric quantities like torsion while relaxing the metricity condition. This formulation encompasses theories such as teleparallel gravity and, as a special case, Weyl conformal  gravity - where the non-metricity tensor plays a fundamental role - opening new perspectives for understanding gravitation in extreme regimes and its potential unification with other fundamental interactions~\cite{bahamonde2023teleparallel,cai2016f,aldrovandi2012teleparallel}.

The proposal to modify GR through gauge symmetries and extensions of its internal structure - such as in MAG - has gained increasing relevance recently. However, applying the gauge principle to gravitation is not a new idea, being primarily inspired by the success of the SM, whose structure is entirely based on this principle. This suggests gauge invariance may serve as a unifying foundation for all fundamental interactions. A central challenge in this approach involves identifying appropriate gauge groups to describe spacetime symmetries. Under certain assumptions, the Coleman-Mandula theorem~\cite{coleman1967all} restricts the possibilities to three bosonic groups: (i) the Poincaré group, (ii) the full conformal group, and (iii) the Weyl group. This work focuses specifically on gravitational theories based on the latter case - Weyl conformal geometry - exploring its implications for a unified description of gravitation under the gauge principle.

Weyl conformal  geometry constitutes a genuine gauge theory, distinguished from other gravitational approaches by its incorporation of a dynamical vector boson - the Weyl boson $\omega_{\mu}$ - introduced through the connection and thus endowed with physical significance~\cite{charap1974gauge,condeescu2025gauge}. This characteristic fundamentally differentiates it from theories based on the Poincaré (i) or full conformal (ii) groups, which lack an analogous dynamical gauge field. The imposition of Weyl symmetry dramatically reduces the number of possible gravitational actions, constraining them to highly specific forms. For instance, in d=4 dimensions without matter or topological terms, the Weyl gravitational action reduces to just two quadratic terms: the square of the Weyl geometry scalar curvature term, $\tilde{R}^2$, and the square of the Weyl tensor $\tilde{C}^2_{\mu\nu\alpha\beta}$, demonstrating the structural rigidity imposed by this symmetry. Although Weyl geometry can be re-expressed in terms of conventional (pseudo-)Riemannian geometry, such reformulation introduces nontrivial modifications: all standard geometric quantities acquire additional contributions dependent on the Weyl boson. Dirac's linearized version~\cite{dirac1973long}, $\phi ^2 \tilde{R}$, which required introducing a compensating scalar field for consistency, proves particularly noteworthy as it reproduces Einstein's theory in the decoupling limit, thereby circumventing historical criticisms of Weyl's gauge approach and revitalizing its potential as a theory of gravitation~\cite{ghilencea2019weyl,jimenez2016cosmology}. The role of Weyl gauge symmetry in this model building beyond SM was studied before~\cite{smolin1979towards,hayashi1977elementary,dirac1973long}. Additionally, the identification of the Weyl gauge field with vector torsion has been proposed in several Refs.~\cite{obukhov1982conformal,klemm2020einstein,barker2025every,condeescu2024weyl}.

Beyond the important role of gauge in these theories, the conformal aspect is also essential. The function of local conformal symmetry in gravity was seriously investigated by Gerard 't Hooft, who promoted it to an exact symmetry~\cite{t2015local}. The scale-dependent effects observed in currently tested energy regimes originate from a low-energy effective theory (general relativity) obtained via spontaneous symmetry breaking. In this approach, the scale factor of the metric in the conformal transformation is interpreted as a dilaton scalar field. Indeed, in gravity, the scarce understanding of the symmetry connecting different scales is becoming increasingly clear. Therefore, the hope is that this possible exact symmetry can help us understand gravity at small scales (such as quantum gravity)~\cite{tHooft:2017avq} or even at cosmological scales~\cite{Bars:2013yba}.

This work investigates Weyl quadratic gravity in the context of black holes. The original Weyl theory, i.e. $\tilde{R}^2$, exhibits a particularly interesting mechanism: it can undergo spontaneous symmetry breaking via the Stueckelberg mechanism, reducing to Einstein gravity while generating an Einstein-Proca theory where the Weyl field acquires mass~\cite{ghilencea2019weyl,ghilencea2020stueckelberg}. The crucial aspect of this process lies in the dynamics of the conformal factor, governed by a scalar field emerging naturally from higher-derivative degrees of freedom, transforming the theory into a scalar-tensor gravitational system. The $\phi(x)$, which we call the
``dilaton", absorbs the scale transformations and is analogous to the dilaton in string theory. Remarkably, this approach dynamically generates both the Planck mass scale and cosmological constant~\cite{ghilencea2019spontaneous}. Recent applications of this theoretical framework have been explored in various contexts including inflation~\cite{ghilencea2019weyl1}, black hole formation and properties~\cite{yang2022black,Becar:2023jtd,Sakti:2024pze}, stellar structure~\cite{Haghani:2023nrm}, the propagation of light in gravitational fields~\cite{Oancea:2023ylb}, galactic dynamics~\cite{Craciun:2023bmu,burikham2023dark}, cosmology~\cite{Harko:2024fnt}, and even particle physics~\cite{de2017local,ghilencea2019weyl,nishino2011weyl}.

Despite extensive study of Weyl conformal  gravity in various contexts, most work has not systematically examined corrections introduced by the square of the Weyl tensor, $\tilde{C}^2_{\mu\nu\alpha\beta}$. While higher-order curvature terms typically present challenges like ghost states, unitarity violations, and instabilities, recent evidence suggests that under certain conditions, theories incorporating this term in a conformal context may exhibit more stable and physically consistent behavior~\cite{mannheim2021solution}. Moreover, solutions for compact objects - particularly static, spherically symmetric black holes - remain unexplored within Weyl conformal  geometry when this term is included. Addressing this gap, our work focuses exclusively on effects of the $\tilde{C}^2_{\mu\nu\alpha\beta}$ term, yielding two principal results: First, we demonstrate that Weyl conformal  gravity admits a black hole solution analogous to the Mannheim-Kazanas case~\cite{mannheim1989exact}, but with the Rindler acceleration term modified by the Weyl gauge coupling constant. Second, we show the theory can undergo symmetry breaking, reducing to an Einstein-Scalar theory with a higher-curvature term that enables new black hole solutions beyond GR predictions.
\section{WEYL GEOMETRIC CONFORMAL GRAVITY}
\subsection{Weyl Conformal Geometry}
The Weyl gauge symmetry or gauged dilatations symmetry is governed by the following transformations~\cite{ghilencea2020stueckelberg}
\begin{equation}\label{weyltrans}
    \hat{g}_{\mu\nu} = \Omega g_{\mu\nu}, \ \ \  \hat{\phi} = \Omega^{-1/2}\phi, \ \ \ \hat{\omega}_{\mu} = \omega_{\mu} -\partial_{\mu}\ln{\Omega}.
\end{equation}
where $\omega_{\mu}$ is called the Weyl vector field or Weyl gauge boson of dilatations and $\phi$ is a scalar field. 
The Weyl geometry can thus be seen as a gauge theory of dilatations.

As rigorously demonstrated in Ref.~\cite{condeescu2025gauge}, Weyl conformal geometry is a dilatation gauge theory. This geometry is defined by its distinctive feature of non-metricity, meaning the covariant derivative of the metric tensor does not vanish, i.e.,
 \begin{equation}\label{nonmetr}
    \tilde{ \nabla}_{\alpha} g_{\mu\nu} = -q \omega_{\alpha}g_{\mu\nu}, \ \ \  \ \ \ \ \ \ \tilde{ \nabla}_{\alpha} g^{\mu\nu} = q \omega_{\alpha}g^{\mu\nu}
 \end{equation}
 where the $q$ is the Weyl gauge coupling constant. Moreover, $ \tilde{ \nabla}_{\alpha}$ is defined
by the Weyl symmetric connection denote $\tilde{\Gamma}^{\alpha}_{\mu\nu}=\tilde{\Gamma}^{\alpha}_{\nu\mu}$, so that the system is torsion-free. Thus, from the non-metricity condition~\eqref{nonmetr}, we directly obtain the Weyl geometry connection, by replacing $\partial_{\mu}\rightarrow  \partial_{\mu} + \omega_{\mu}$, given by
\begin{equation}\label{conecweyl}
    \tilde{\Gamma}^{\alpha}_{\mu\nu} =    \Gamma^{\alpha}_{\mu\nu} +     \Xi^{\alpha}_{\mu\nu},
\end{equation}
where
\begin{equation}
    \Xi^{\alpha}_{\mu\nu} =  \frac{q}{2} \bigg[\delta^{\alpha} _{\mu}\omega_{\nu}  +\delta^{\alpha} _{\nu}\omega_{\mu}-g_{\mu\nu}\omega^{\alpha}  \bigg].
\end{equation}
The $\Gamma^{\alpha}_{\mu\nu}$
is the  standard
Levi-Civita connection, associated with the metric $g_{\mu\nu}$, i.e., $\nabla_{\alpha}g_{\mu\nu}=0$, where $\nabla_{\alpha}$ is defined by the Levi-Civita connection.


From the connection~\eqref{conecweyl}, 
it is possible to obtain the curvature tensor. The Riemann tensor in Weyl conformal geometry is given by
\begin{equation} \label{riemweyl}
    \tilde{R}^{\alpha}_{\mu\nu\lambda}= \partial_{\nu}\tilde{\Gamma}^{\alpha}_{\mu\lambda}-\partial_{\lambda}\tilde{\Gamma}^{\alpha}_{\mu\nu}+\tilde{\Gamma}^{\alpha}_{\sigma\nu}\tilde{\Gamma}^{\sigma}_{\mu\lambda}-\tilde{\Gamma}^{\alpha}_{\sigma\lambda}\tilde{\Gamma}^{\sigma}_{\mu\nu}.
\end{equation}
 So that the Ricci tensor and Ricci scalar are given by, respectively, $\tilde{R}_{\mu\nu} = \tilde{R}^{\alpha}_{\mu\alpha\nu}$ and $\tilde{R}=g^{\mu\nu}\tilde{R}_{\mu\nu}$. These quantities can be expressed in terms of standard GR quantities (written with the Levi-Civita connection) plus corrections introduced by Weyl geometry, i.e., in terms of the Weyl field. By substituting the connection~\eqref{conecweyl} into the Eq.~\eqref{riemweyl}, we find that the Riemann tensor in this post-Riemannian geometry is given by
 \begin{equation}
     \tilde{R}^{\alpha}_{\mu\nu\lambda} = R^{\alpha}_{\mu\nu\lambda}+\nabla_{\nu}\Xi_{\mu\lambda}^\alpha-\nabla_{\lambda}\Xi_{\mu\nu}^\alpha+ \Xi_{\sigma\nu}^\alpha\Xi_{\mu\lambda}^\sigma-\Xi_{\sigma\lambda}^\alpha\Xi_{\mu\nu}^\sigma
 \end{equation}
 where $R^{\alpha}_{\mu\nu\lambda}$ is the curvature tensor of Riemann geometry.  The scalar Ricci is given by
 \begin{equation}
     \tilde{R} = R - 3 q \nabla_{\mu}\omega^{\mu}-\frac{3}{2}q^2\omega_{\mu}\omega^{\mu},
 \end{equation}
where $R$ denotes the Ricci scalar, defined in the Riemann
geometry. Another fundamental quantity for our analysis is the Weyl tensor $    \tilde{C}_{\mu\nu\alpha\beta}$, which plays a central role in conformally invariant gravitational theories~\cite{mannheim2012making}. The quadratic invariant of the Weyl tensor takes the following simple form:
\begin{equation} \label{tensorweyl1}
    \tilde{C}^2_{\mu\nu\alpha\beta} = {C}^2_{\mu\nu\alpha\beta} + \frac{3}{2}q^2 \tilde{F}_{\mu\nu}^2
\end{equation}
where the ${C}_{\mu\nu\alpha\beta}$  is the Weyl tensor\footnote{The Weyl tensor in $d=4$ is defined by
\begin{align}\nonumber
&C_{\mu\nu\alpha\beta} = R_{\mu\nu\alpha\beta} 
- \left( g_{\mu\alpha} R_{\nu\beta} - g_{\mu\beta} R_{\nu\alpha} - g_{\nu\alpha} R_{\mu\beta} + g_{\nu\beta} R_{\mu\alpha} \right) \\ \nonumber
&+ \frac{R}{3} \left( g_{\mu\alpha} g_{\nu\beta} - g_{\mu\beta} g_{\nu\alpha} \right)
\end{align}}, defined in the standard
Riemannian geometry and the field strength of the Weyl gauge  field is defined by.
\begin{equation}
    \tilde{F}_{\mu\nu} = \partial_{\mu}\omega_{\nu}-\partial_{\nu}\omega_{\mu}.
\end{equation}
Additionally, we can represent ${C}^2_{\mu\nu\alpha\beta}$ as an expansion in Ricci tensors. Applying the Gauss identity, i.e., $\mathcal{G} = R^2-4R_{\mu\nu}R^{\mu\nu}+R_{\mu\nu\alpha\beta}R^{\mu\nu\alpha\beta}$ , yields
\begin{equation}\label{tensorweyl2}
  {C}^2_{\mu\nu\alpha\beta} = -\frac{2}{3}R^2 +2 R_{\mu\nu}R^{\mu\nu} + \mathcal{G}. 
\end{equation}
Key properties of these post-Riemannian tensors are derived in Ref.~\cite{oancea2024weyl}. Notably, the symmetries of Riemann tensor in Weyl geometry reduces to its Riemannian counterpart only if $\tilde{F}_{\mu\nu}=0$, i.e., the $\omega_{\mu}$ is a gradient vector field. For this gauge choice, we refer to it as Weyl integrable geometry~\cite{salim1998spherically,salim1999gravitational,scholz2016clusters}.

\subsection{Action and field equations}
 The most general action in Weyl geometry that is invariant under gauged~\cite{weyl1919new} dilatations~\eqref{weyltrans} is given by
 \begin{equation}\label{ac}
    S =  \int d^4 x \sqrt{-g}(\alpha_1 \tilde{R}^2+\alpha_2 \tilde{F}_{\mu\nu}^2 +\alpha_3 \tilde{C}_{\mu\nu\lambda\sigma} \tilde{C}^{\mu\nu\lambda\sigma} ),
\end{equation}
where $\alpha_{1,2,3}$ are dimensionless coefficients. The first two terms in the action above have been extensively studied in the literature. For instance, Ref.~\cite{ghilencea2019weyl,ghilencea2020stueckelberg,ghilencea2019spontaneous} demonstrated that this Weyl-geometry-inspired quadratic action exhibits spontaneous symmetry breaking via the Stueckelberg mechanism into Einstein-Proca action, leading to the key result that the Weyl gauge field acquires mass $m_{\omega}$. So that after $\omega_{\mu}$ decouples, below $m_{\omega}$ the Einstein-
Hilbert action is obtained as a ‘low-energy’ effective theory
of Weyl gravity. At this point, the change in the transition between symmetries happens dynamically at the level of the Lagrangian, whose 'gauge fixing' is done through a transformation $    \Omega = \frac{\phi^2}{<\phi^2>}$,  that is scale-dependent, in which $<\phi^2>$ is fixed at the vacuum expectation value (VEV).

However, as previously mentioned, higher-derivative theories, such as the original Weyl gravity, are known to host ghosts in their gravitational spectrum. Although it has been demonstrated that this pathological feature can be circumvented under certain conditions via a symmetry-breaking mechanism, a significant gap remains in the literature. Specifically, a more comprehensive analysis of the propagating gravitational modes utilizing the Weyl tensor squared term is notably absent. This particular term, which introduces a massive spin-2 ghost mode, is of paramount importance in the quantum level, when trying to renormalize SM in the
presence of gravity~\cite{hooft2017local}.


In a first approach, we investigate exclusively the corrections introduced by the Weyl tensor squared term to black hole solutions. Although the theory is potentially not free from pathological degrees of freedom (see Ref. for a contrary viewpoint), we adopt an effective approach to examine the effects of this theory within the classical regime. Furthermore, it is important to emphasize that a detailed analysis of the gravitational spectrum of action~\eqref{ac} remains absent. In light of the above, we adopt the following action as the foundation for this work:
\begin{equation}\label{ac1}
    S = 2\alpha^2 \int d^4 x \sqrt{-g} \tilde{C}_{\mu\nu\lambda\sigma} \tilde{C}^{\mu\nu\lambda\sigma} 
\end{equation}
where $\alpha$ is a dimensionless constant. Substituting Eqs.~\eqref{tensorweyl1} and~\eqref{tensorweyl2} into the above action, we find the following post-Riemannian conformally invariant quadratic action (omitting the topological term $\mathcal{G}$ in $d=4$ dimensions):
\begin{equation} \label{aceff}
    S = 2\alpha^2 \int d^4 x \sqrt{-g} \bigg[  -\frac{2}{3}R^2 +2 R_{\mu\nu}R^{\mu\nu} + \frac{3}{2}q^2 \tilde{F}_{\mu\nu}^2 \bigg]. 
\end{equation}
Note that each term above is Weyl gauge invariant. Moreover, the effective action~\eqref{aceff} is essentially equivalent to that employed in the works of Mannheim and Kazanas, but incorporates a residual term from Weyl geometry - specifically, the kinetic term of the Weyl vector boson.

To derive the gravitational equations of motion, we vary action~\eqref{ac} with respect to the metric, which gives us the simple result:
\begin{equation}\label{weyleom}
 B_{\mu\nu} = \frac{3}{2}q^2 T_{\mu\nu}^{\omega}
\end{equation}
where $B_{\mu\nu}$ is the Bach tensor and $T_{\mu\nu}^{\omega}$ is the stress-energy tensor of Weyl field, given by, respectively,
\begin{align}
    B_{\mu\nu} &\equiv 2 \nabla^\rho \nabla^\sigma C_{\mu\rho\nu\sigma} + C_{\mu\rho\nu\sigma} R^{\rho\sigma} , \\
    T_{\mu\nu}^{\omega} &\equiv \tilde{F}_{\mu\rho} \tilde{F}_\nu^{\ \rho} - \frac{1}{4} g_{\mu\nu} \tilde{F}_{\alpha\beta} \tilde{F}^{\alpha\beta}.
\end{align}
The $ B_{\mu\nu}$ has up to fourth-order derivatives, making~\eqref{weyleom}  a fourth-order theory.

The next step involves introducing an auxiliary scalar field $\phi$ to extract the additional degree of freedom contained in the invariant $R^2$. To achieve this, we perform the replacement $R^2 \rightarrow -2 \phi^2 R -\phi^4$. Note that it is straightforward to verify the mathematical equivalence between the new and original actions. This result follows from using the solution 
\begin{equation}\label{phi}
\phi^2 = -R
\end{equation}
obtained from the equation of motion for $\phi$ - an approach widely adopted in the literature.  This field $\phi$ would thus correspond to the dilatonic Goldstone mode—a spin-zero state propagated by the $R^2$ term in the action—which, as will be shown later, is responsible for spontaneous symmetry breaking. Thus, the new action we will investigate is given by
\begin{equation}\label{actionboson}
    S = 2\alpha^2 \int d^4 x \sqrt{-g} \bigg[ \frac{4}{3}\phi^2R +\frac{2}{3}\phi^4 +2 R_{\mu\nu}R^{\mu\nu} + \frac{3}{2}q^2 \tilde{F}_{\mu\nu}^2 \bigg]  
\end{equation}

By varying the action Eq.~\eqref{actionboson} with respect to the metric tensor we find the gravitational field equations of the Weyl geometric gravitational theory,
\begin{align}\label{eq11}
    &\phi^2G_{\mu\nu}-\frac{\phi^4}{4}g_{\mu\nu} -2 (\nabla_{\mu}\phi\nabla_\nu\phi+\phi\nabla_\mu\nabla_{\nu}\phi) \\ \nonumber
    &+2g_{\mu\nu}(\phi \Box \phi +\nabla_{\lambda}\phi\nabla^{\lambda}\phi) - \frac{9}{4}q^2(-\tilde{F}_{\mu}{}^{\lambda}\tilde{F}_{\nu\lambda}\\ \nonumber
    &+\frac{g_{\mu\nu}}{4}\tilde{F}_{\lambda\sigma}\tilde{F}^{\lambda\sigma})+\frac{3}{2}(R^{\lambda\sigma}R_{\mu\lambda\nu\sigma}-\nabla_{\mu}\nabla_{\nu}R + \Box R_{\mu\nu}) \\ \nonumber
    &+\frac{3}{4}g_{\mu\nu}(-R_{\lambda\sigma}R^{\lambda\sigma}+\Box R)=0.
\end{align}
The trace of Eq.~\eqref{eq11} gives
\begin{equation}
    \phi^2R + \phi^4 = 3 \Box (R+\phi^2 )
\end{equation}
Thus, as expected, the constraint $\phi^2=-R$ derived from the equation of motion of the scalar field satisfies the above equation. On the other hand, the variation of Eq.~\eqref{actionboson} with respect to $\omega_{\mu}$ gives the equation of motion of the Weyl vector as
\begin{equation}\label{eF}
    \nabla_{\mu}\tilde{F}^{\mu\nu}=\frac{1}{\sqrt{-g}}\partial_{\mu}(\sqrt{-g}\tilde{F}^{\mu\nu})=0.
\end{equation}
\section{THE SPHERICALLY SYMMETRIC
VACUUM FIELD EQUATIONS}
In order to examine the influence of Weyl geometry on static, spherically symmetric vacuum solutions for compact objects, we employ the following metric ansatz:
\begin{equation} \label{metric}
    ds^2=- A(r)dt^2+\frac{dr^2}{A(r)}+r^2(d\theta^2+\sin^2\theta d\phi^2).
\end{equation}
Considering a spacetime characterized by the metric~\eqref{metric} and utilizing the gauge freedom inherent to the Weyl vector field, we find that the most general Weyl strength field (2-form) admits the form: 
\begin{equation}\label{F}
    \tilde{\textbf{F}}_2 = \omega_0(r) dt \wedge dr + \tilde{P}\sin\theta d\theta \wedge d\phi,
\end{equation}
where the constant $\tilde{P}$ plays the role of a Weyl magnetic charge. Substituting the Eq.~\eqref{F} into Eq.~\eqref{eF}, we obtain that
\begin{equation}
    \omega_0 = - \frac{\tilde{Q}}{r},
\end{equation}
where $\tilde{Q}$ plays the role of a Weyl electric charge.

By inserting the ansatz~\eqref{metric} and Eq.~\eqref{F} into Eq.~\eqref{weyleom}, we derive the higher-order gravitational field equations governing the metric function $A(r)$, which take the form 
\begin{align}\nonumber
 & \frac{2 A \left(2 A'+r \left(r \left(A^{(4)} r+3 A^{(3)}\right)-A''\right)\right)}{3 r^3}+\frac{1}{3} A^{(3)} A' \\ 
  & -\frac{\left(r A''-2 A'\right)^2}{6 r^2} - \frac{4 A^2-9 q^2 \left(\tilde{P}^2+\tilde{Q}^2\right)-4}{6 r^4}=0 \label{eq1}
\end{align}
\begin{align}\nonumber
  &  \frac{A^{(3)} \left(r A'-2 A\right)}{3 r} -\frac{\left(r \left(r A''-2 A'\right)+2 A\right)^2}{6 r^4} \\
  & + \frac{9 q^2 \left(\tilde{P}^2+\tilde{Q}^2\right)+4}{6 r^4}=0\label{eq2}
\end{align}
\begin{align}\nonumber
  &  -\frac{A \left(4 A'+r \left(r \left(A^{(4)} r+2 A^{(3)}\right)-2 A''\right)\right)}{3 r^3}-\frac{1}{3} A^{(3)} A'\\ 
  &+\frac{\left(r A''-2 A'\right)^2}{6 r^2}+\frac{4 A^2-9 q^2 \left(\tilde{P}^2+\tilde{Q}^2\right)-4}{6 r^4}=0\label{eq3}.
\end{align}
Subtracting the Eq~\eqref{eq1} of Eq.~\eqref{eq2}, we obtain that
\begin{equation}\label{A}
    \frac{2 A \left(A^{(4)} r+4 A^{(3)}\right)}{3 r}=0.
\end{equation}
The solution to Eq.~\eqref{A} yields:
\begin{equation}\label{MK}
A(r) = a_1 + \frac{a_2}{r} + a_3 r + a_4 r^2,
\end{equation}
where $a_i$ ($i=1,\dots,4$) are integration constants. By substituting this ansatz into either Eq.~\eqref{eq1}, \eqref{eq2}, or \eqref{eq3}, we obtain the following consistency condition:
\begin{equation}
a_3 = \frac{4(a_1^2 - 1) - \tilde{q}^2}{12a_2},
\label{eq:constraint}
\end{equation}
where $\tilde{q}^2 \equiv 9q^2(\tilde{P}^2 + \tilde{Q}^2)$. This constraint is necessary for the solution~\eqref{MK} to satisfy the complete set of gravitational field equations. Note that when $\tilde{q}=0$, we recover the well-know MK solution. 

It is important to emphasize that despite the presence of a vector field (Weyl) with dynamics similar to Maxwell's, the obtained solution~\eqref{MK} does not exhibit a Reissner-Nordström-like term generated by the vector boson. As we will show later, this term only emerges when the Weyl conformal symmetry is spontaneously broken.

The line element can be be written as 
\begin{align} \label{metric1} \nonumber
   & ds^2=- \bigg(a_1 + \frac{a_2}{r} + a_3 r + a_4 r^2\bigg)dt^2 \\
    &+\frac{dr^2}{\bigg(a_1 + \frac{a_2}{r} + a_3 r + a_4 r^2\bigg)}+r^2(d\theta^2+\sin^2\theta d\phi^2).
\end{align}
where the Rindler acceleration term is given by
\begin{equation}
  a_3 = \frac{4(a_1^2 - 1) - \tilde{q}^2}{12a_2}
\end{equation}
We highlight here the role of non-metricity, characterized by the Weyl gauge coupling constant. Unlike the previously mentioned solutions which have the form~\eqref{MK}, the Rindler acceleration term in our current approach can be interpreted through the new vector degree of freedom introduced by Weyl conformal geometry. It is worth emphasizing that the presence of this linear term in $r$ in the metric~\eqref{metric1} has been used, within various modified gravity theories~\cite{yang2022black,mannheim1989exact,panpanich2018fitting}, to explain the flat rotation curves of galaxies without invoking dark matter.

From Eq.~\eqref{phi}, we obtain that
\begin{equation} \label{scalar}
    \phi = \sqrt{\frac{2(a_1-1)}{r^2}+\frac{6 a_3}{r}+12 a_4}.
\end{equation}
Note that the scalar field approaches a constant value asymptotically at infinity. Notably, it has been shown in a minimal extension of the standard model that a $SU(2)\times U(1)$ singlet scalar field $\phi(x)$ can be a viable dark matter candidate~\cite{bars2006standard}. This is because the gauge symmetry prevents this field from coupling to ordinary matter fields.

Note that solution~\eqref{metric1} features three free integration constants. If we set $a_3 = 0$, we recover a family of Schwarzschild-de Sitter solutions. In this case, if we ``switch off" the Weyl gauge geometry (i.e., $\tilde{q} = 0$), we consequently find that $a_1 = 1$. However, this family of solutions is not only valid in Weyl conformal  gauge gravity but also in General Relativity. It is within the framework of GR that we can safely and unambiguously fix these constants. Therefore, applying a mechanism to recover the Einstein-Hilbert action is highly desirable, as it provides a secure foundation for interpreting these parameters. In the next section, we will employ a mechanism of spontaneous symmetry breaking to achieve this objective

\section{Spontaneous breaking of Weyl gauge symmetry}
The physical world we inhabit is scale-dependent, in contrast to the realm of high-energy scales—or short distances—where fundamental interactions are expected to behave differently than at low energies. As we have seen, the results derived from our action~\eqref{ac1} are scale-independent, since the model is conformally invariant (Weyl gauge symmetry). Therefore, if we aim to describe gravitational compact objects within low-energy models, a mechanism is required to explain how these two distinct regimes can be connected. As previously proposed, this can be achieved through spontaneous symmetry breaking. The key to this mechanism is to promote the scalar field $\phi$ to the conformal factor $\Omega$. To demonstrate this, we apply a specific form of transformation~\eqref{weyltrans} that is scale-dependent $    \Omega = \frac{\phi^2}{<\phi^2>}$, i.e., $\phi$ is fixed to its VEV. If we assume that $<\phi^2> = \frac{3 M_{Pl}^2}{16 \alpha^2}$, i.e., 
\begin{equation}\label{gauge}
    \Omega = \frac{16 \alpha^2}{3 M_{Pl}^2}\phi^2,
\end{equation}
we obtain from action~\eqref{actionboson} the following effective action in the Einstein frame:
\begin{align}\label{actioneffec} \nonumber
   & S =  \int d^4 x \sqrt{-\hat{g}} \bigg[ \frac{M_{Pl}^2}{2}\hat{R} + \frac{3 M_{Pl}^4}{128\alpha^2} +4\alpha^2 \hat{R}_{\mu\nu}\hat{R}^{\mu\nu} \\
   &+ 3\alpha^2q^2\hat{F}_{\mu\nu}^2 + \mathcal{L}_{\phi}^{M_{Pl}}+\mathcal{L}_{\phi}^{\alpha} \bigg]
\end{align}
\begin{align}\label{phim}
    \mathcal{L}_{\phi}^{M_{Pl}} =\frac{M_{Pl}^2}{2}\bigg(3\frac{\hat{\Box}\phi^2}{\phi^2}-\frac{3}{2}\frac{\hat{\nabla}_{\mu}\phi^2\hat{\nabla}^{\mu}\phi^2}{\phi^4}\bigg),
\end{align}
\begin{align}\nonumber \label{phia}
&\mathcal{L}_{\phi}^{\alpha} =4\alpha^2 \bigg( 2\frac{\hat{R}^{\mu\nu}\hat{\nabla}_{\mu}\hat{\nabla}_{\nu}\phi^2}{\phi^2}  + \hat{R}\frac{\hat{\Box}\phi^2}{\phi^2}+3\frac{(\hat{\Box}\phi^2)^2}{\phi^4}   \\\nonumber
    &-3\frac{\hat{\Box}\phi^2 \hat{\nabla}_{\mu}\phi^2\hat{\nabla}^{\mu}\phi^2}{\phi^6}  +\frac{3}{4}\frac{(\hat{\nabla}_{\mu}\phi^2\hat{\nabla}^{\mu}\phi^2)^2}{\phi^4}  \\
    &-\frac{\hat{\nabla}^{\mu}\hat{\nabla}^{\nu}\phi^2\hat{\nabla}_{\mu}\phi^2\hat{\nabla}_{\nu}\phi^2}{\phi^6}\bigg).
\end{align}
That is, a Weyl gauge fixing symmetry transformation applied to Weyl quadratic gravity~\eqref{ac} without matter yields an Einstein-Higher-curvature-Horndeski action with a positive cosmological constant term for the Weyl gauge field.

The next step is to find the static, spherically symmetric vacuum solution for the action~\eqref{actioneffec} in the Einstein frame. Using gauge condition~\eqref{gauge} and solution~\eqref{scalar} through the conformal transformation~\eqref{weyltrans}, we obtain
\begin{equation} \label{completa}
    ds^2 = \hat{g}_{\mu\nu}dx^{\mu}dx^{\nu}= \bigg(\frac{16 \alpha^2}{3 M_{Pl}^2}\phi^2\bigg) g_{\mu\nu}dx^{\mu}dx^{\nu}.
\end{equation}

\begin{align} \label{g00}
    \hat{g}_{00} = - \frac{16 \alpha^2}{3 M_{Pl}^2}\left(\frac{2(a_1-1)}{r^2}+\frac{6 a_3}{r}+12 a_4\right) \\
    \times \left(a_1 + \frac{a_2}{r} + a_3 r + a_4 r^2 \right)
\end{align}
\begin{equation}\label{g11}
    \hat{g}_{11} = \frac{16 \alpha^2}{3 M_{Pl}^2}\frac{\left(\frac{2(a_1-1)}{r^2}+\frac{6 a_3}{r}+12 a_4\right)}{\left(a_1 + \frac{a_2}{r} + a_3 r + a_4 r^2 \right)} 
\end{equation}
\begin{equation}
    \hat{g}_{22} = \frac{16 \alpha^2}{3 M_{Pl}^2}\left(\frac{2(a_1-1)}{r^2}+\frac{6 a_3}{r}+12 a_4\right) r^2
\end{equation}
where the coefficient $a_3$ is given by Eq.~\eqref{eq:constraint}.

In order to express the line element in the Einstein frame in a more conventional form, we adopt the following coordinate transformation: $\hat{g}_{22} = x^2$,which relates the old coordinate to the new one via
\begin{equation}\label{coorde}
    r =-\frac{a_3}{4 a_4}+\sqrt{\frac{1-a_1}{6 a_4}+\frac{a_3^2}{16 a_4^2}+\frac{M_{Pl}^2}{64 a_4 \alpha ^2 }x^2}.
\end{equation}
Substituting the transformation~\eqref{coorde} into Eqs.~\eqref{g00} and~\eqref{g11}  and correctly computing the $dx^2$ term, we obtain the new solution in the Einstein frame, expressed in the $x$ coordinate, as follows:
\begin{equation}
    ds^2= -\tilde{A}(x)d\tilde{t}^2+\frac{dx^2}{\tilde{B}(x)}+x^2d\Omega^2
\end{equation}
where $d\tilde{t}^2= \frac{64a_4\alpha^2}{M_{Pl}^2}dt^2$. The complete expressions for the  metric functions $\tilde{A}$ and $\tilde{B}$ are in the appendix~\eqref{apen}. In limit that $a_3=0$ and $a_1=1$, we obtain
\begin{equation} \label{sch}
   \tilde{A}(x)=\tilde{B}(x)^{-1} = 1 +\frac{8  a_2 \sqrt{a_4}\alpha}{M_{Pl} x} +\frac{M_{Pl}^2 x^2}{64 \alpha^2}. 
\end{equation}
Set $\frac{8  a_2 \sqrt{a_4}\alpha}{M_{Pl} }=-R_{sc}$, where $R_{sc}$ is radius of Schwarzchild, thus we find that $   a_4 = \frac{R_{sc}^2 M_{Pl} ^2}{64 \alpha^2 a_2^2}$. If we assume that $a_2=- R_{Sc}$, so $ a_4 = \frac{ M_{Pl} ^2}{64 \alpha^2}$. It should be noted that, in this limit ($a_1=1$ and $a_3=0$), the Schwarzschild-de Sitter metric~\eqref{sch} satisfies not only the Einstein-Hilbert action but also an extended action incorporating a quadratic curvature term. So that the Weyl line element~\eqref{metric1} is given by 
\begin{align} \label{metric2} \nonumber
   & ds^2=- \bigg(1 - \frac{R_{Sc}}{r} + \frac{\tilde{q}^2}{12R_{Sc}} r + \frac{M_{Pl}^2}{64\alpha^2} r^2\bigg)dt^2 \\
    &+\frac{dr^2}{\bigg(1 - \frac{R_{Sc}}{r} + \frac{\tilde{q}^2}{12R_{Sc}} r + \frac{M_{Pl}^2}{64\alpha^2} r^2\bigg)}+r^2(d\theta^2+\sin^2\theta d\phi^2).
\end{align}

\begin{figure}[!ht] 
\includegraphics[height=5cm]{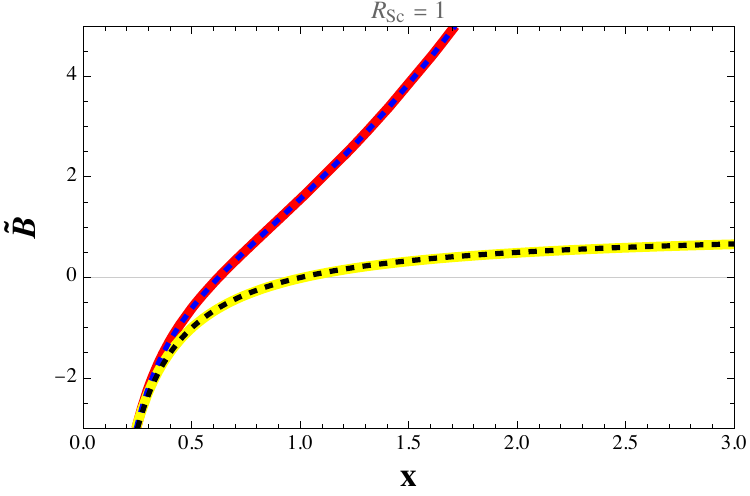}
\caption{Horizons with small $\tilde{q}$.}  \label{hor1}
\end{figure}

\begin{figure}[!ht] 
\includegraphics[height=5cm]{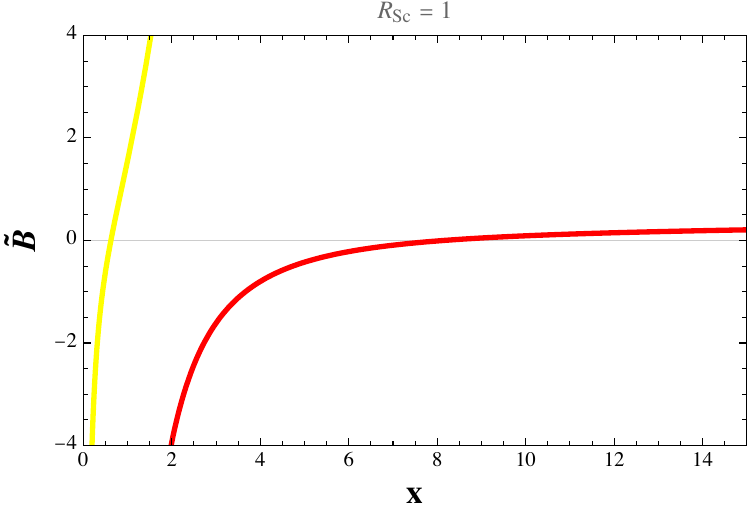}
\caption{Horizons with large $\tilde{q}$.}  \label{hor2}
\end{figure}

Due to the complexity of the expressions for $\tilde{A}$ and $\tilde{B}$ (see the Appendix~\ref{apen}), we investigate the solutions in two regimes of interest: (i) when the high-curvature terms are suppressed by the Planck mass ($\epsilon\gg 1$) , and (ii) the opposite case, where higher-derivative corrections become significant ($\epsilon\ll1$); where $\epsilon = \frac{M_{Pl}}{\alpha}$. Before proceeding, it is possible to investigate some geometric features of these solutions, such as event horizons, which are located at $B(x)=0$ in Eq.~\eqref{Bmodif}. Specifically, in Fig.~\ref{hor1}, we show the location of the horizons under the assumption of a small $\tilde{q}$. The solid red and yellow lines represent the residual effects introduced by Weyl non-metricity for scenarios where $\epsilon \gg 1$ and $\epsilon \ll 1$, respectively. The dashed blue and black lines correspond to the respective cases where the effects of residual  Weyl dilation are disregarded. Note that, due to the choice of a small $\tilde{q}$, the spontaneous breaking of Weyl geometry causes little interference with the location of the horizons. However, the effects become observable when we consider higher-derivative corrections compared to the case where the Einstein-Hilbert term is dominant. On the other hand, for a large $\tilde{q}$ in Fig.~\ref{hor2}, we note not only changes in the horizon locations for the different regimes but also a change in divergence behavior at the origin: for the case $\epsilon \ll 1$, the solution tends to improve the central singularity.

\begin{itemize}
    \item \textbf{Assuming $M_{Pl}\gg\alpha$ or $\epsilon\gg1$}: 
\end{itemize}
The first regime of interest occurs when the Planck mass suppresses the higher-derivative term, i.e., $\epsilon\gg1$ . In this case, the dominant gravitational contributions originate solely from the Einstein-Hilbert term, supplemented by scalar field corrections arising from the Lagrangian in Eq.~\eqref{phim}. Under the assumption that $\epsilon \gg 1$, the transformation in Eq.~\eqref{coorde} for $a_1=1$, $a_2= -R_{Sc}$, $a_3= \frac{\tilde{q}^2}{12R_{Sc}}$ and $a_4=\frac{M_{Pl}^2}{64\alpha^2}$ is approximately given by
\begin{align} \nonumber
   & r \approx x -\frac{4\tilde{q}^2}{3  R_{Sc} \epsilon^2 }.
\end{align}
The metric components~\eqref{Amodif} and~\eqref{Bmodif} for $r_{+}$ are given by
\begin{align}\nonumber
    &\tilde{A}(x) \approx 1 + \frac{\tilde{q}^4}{9 R_{Sc}^2 \epsilon ^2} - \frac{\tilde{R}_{Sc}}{x} + \frac{\tilde{Q}^2}{x^2}+  \frac{\tilde{q}^2 }{12 R_{Sc}} x \\
    &+ \frac{\epsilon^2}{64}x^2 + O(1/\epsilon)^3,
\end{align}
\begin{equation}
    \tilde{B}(x) \approx  1-\frac{\tilde{q}^4}{12 R_{Sc}^2 \epsilon ^2} - \frac{\bar{R}_{Sc}}{x} + \frac{\bar{Q}^2}{x^2} + \frac{\epsilon^2}{64}x^2 + O(1/\epsilon)^3
\end{equation}
where $\tilde{Q}^2=-\frac{4 \tilde{q}^2}{\epsilon ^2}$, $ \bar{Q}^2=\frac{4 \tilde{q}^2}{3 \epsilon ^2}$, 
\begin{equation}
    \tilde{R}_{Sc}=R_{Sc} -\frac{8 \tilde{q}^2}{3 R_{Sc} \epsilon ^2},
\end{equation}
and
\begin{equation}
    \bar{R}_{Sc}=R_{Sc}+\frac{8 \tilde{q}^2}{3 R_{Sc} \epsilon ^2}.
\end{equation}

Note that even in the regime where the high-curvature effects are suppressed, a novel solution persists for which $A(x) \neq B(x)$. Furthermore, as expected, a Reissner-Nordström-like term emerges, proportional to the square of the Weyl gauge coupling constant. Additionally, a magnetic monopole-like term appears, proportional to $\tilde{q}^4$. The cosmological constant term remains unchanged from its form prior to symmetry breaking. It is particularly noteworthy that the Rindler acceleration term not only persists in this regime but also manifests exclusively in the temporal component of the metric! Furthermore, by computing the Kretschmann scalar $K$ in this regime, i.e.,
\begin{equation}
    K_{\epsilon\gg1} \approx \frac{3 \epsilon ^4}{512} +\frac{12 R_{Sc}^2}{x^6}-\frac{\tilde{q}^2}{2 x^4}+\frac{\tilde{q}^4}{32 R_{Sc}^2 x^2}-\frac{\tilde{q}^2 \epsilon ^2}{64 R_{Sc} x}+O(1/\epsilon)^2,
\end{equation}
we conclude that the residual effects of Weyl symmetry do not soften the divergence at the origin.

\begin{itemize}
    \item \textbf{Assuming $M_{Pl}\ll\alpha$ or $\epsilon\ll1$}: 
\end{itemize}
The other energy regime of interest occurs when the effects of high curvature become significant. In this case, we must employ the approximation $\epsilon \ll 1$. Substituting this condition into Eq.~\eqref{coorde}, we find that:
\begin{equation}
   r\approx \frac{3 R_{Sc}  \epsilon ^2}{8  \tilde{q}^2}x^2.
\end{equation}
The metric components~\eqref{Amodif} and~\eqref{Bmodif}  are given by
\begin{align} \nonumber
   & \tilde{A}(x) \approx-\frac{512 \tilde{q}^6}{27 R_{Sc}^2 x^4 \epsilon ^6} + \frac{64 \tilde{q}^4}{9 R_{Sc}^2 x^2 \epsilon ^4}+\frac{2 \tilde{q}^4}{9 R_{Sc}^2 \epsilon ^2}\\ 
    &-\frac{8 \tilde{q}^2}{x^2 \epsilon ^2}+2 + \frac{3 x^2 \epsilon ^2}{64}+ O(\epsilon)^3
\end{align}
\begin{equation}
    \tilde{B}(x)\approx -\frac{2 \tilde{q}^2}{3 x^2 \epsilon ^2}+\frac{1}{4}+\frac{x^2 \epsilon ^2}{128}-\frac{9 R_{Sc}^2 \epsilon ^2}{32 \tilde{q}^2} + O(\epsilon)^3
\end{equation}
Once again, we find that in this approximation $\tilde{A}(x) \neq \tilde{B}(x)$. As expected, we demonstrate that the higher-derivative terms indeed improve the divergence at the origin, as the leading term in the temporal metric component now falls off as $1/x^6$. This behavior can be confirmed by examining the leading-order term of the Kretschmann scalar in this regime, which is approximately given by $K_{\epsilon\ll1}\approx\frac{400 \tilde{q}^4}{3 x^8 \epsilon ^4}$.

\section{Conclusion}\label{con}
In this work, we investigated the effects of the  Weyl conformal gauge symmetry on black hole solutions. Interestingly, Weyl geometry is not only crucial in the early universe or high-energy regimes, as widely argued, but we also demonstrated that signatures of this symmetry can be found in low-energy regimes.

We adopted a conformal quadratic action with the Weyl tensor within the Weyl geometry. Subsequently, we transition to the Riemannian formulation. For this choice, the dynamic of the Weyl gauge field emerges naturally. This new residual vector degree of freedom, originating from the non-metricity condition, introduces novel effects on the geometry of compact objects.

We assumed a static, spherically symmetric vacuum metric with a dyon-like configuration for the Weyl bosonic field. Although the equations of motion are of fourth order, they simplify remarkably in vacuum. We thus find a Mannheim-Kazanas(MK)-type black hole solution in Weyl gauge geometry. In this new solution, we showed that the coefficient of the Rindler acceleration term, $a_3$, depends on the Weyl gauge coupling constant. This allows for a geometric interpretation of this term—unpredicted by General Relativity— as a result of non-metricity. Indeed, by setting $a_3=0$ (which we find implies $q=0$ only if $a_1=1$), we recovered a family of Schwarzschild-de Sitter solutions.

However, it is noteworthy that we recover scale-dependent General Relativity in a specific limit. To achieve this, we utilized the vacuum expectation value (VEV) of an auxiliary scalar field as a gauge via a conformal factor $\Omega$. That is, in the Weyl formulation, a simple Weyl symmetry transformation (a "gauge fixing") applied to an action in Weyl geometry results in an action directly in Riemannian geometry with spontaneous breaking of the Weyl gauge symmetry. In our case,  Weyl quadratic gravity was gauge-transformed into an Einstein-Horndeski action with a high-curvature term and a positive cosmological constant, while still preserving the dynamical Weyl vector field (which remains massless).

From this spontaneous symmetry breaking, we obtain a new black hole solution that is no longer conformally invariant. This new solution exhibits behavior markedly different from the MK-type solution. Initially, we found that in the limit where we disregard the residual effects of Weyl non-metricity ($a_1=1$ and $a_3=0$, consequently $q=0$), we once again recovered the Schwarzschild-de Sitter solution. This finally allowed us to fix the other constants ($a_2$ and $a_4$) that appear in the initial conformal solution. Specifically, we found the cosmological constant term ($a_4$) is given in terms of the Planck mass and the conformal coupling constant $\alpha$, while $a_2$ is the Schwarzschild radius itself. Furthermore, the coefficient of the Rindler acceleration term is given by $\tilde{q}/R_{Sch}$. Therefore, our theory features only two free parameters: $\tilde{q}$ and $\alpha$.

We also studied the approximate behavior of the new solution in different regimes. In the regime where the Planck mass suppresses the high-curvature term ($M_{Pl} / \alpha \gg 1$), we find that even in a low-energy regime, a Rindler acceleration term persists in the temporal component of the metric. This is accompanied by corrections to the Schwarzschild radius proportional to $\tilde{q}^2$ and a Reissner-Nordström-like term generated purely by the Weyl pseudo-charge. On the other hand, we also examine the regime where higher-derivative corrections are significant ($M_{Pl} / \alpha \ll 1$). In this case, we discovered that the divergence at the origin is enhanced by the residual effects of Weyl gauge geometry.

\begin{acknowledgments}
The authors would like to acknowledge Fundação de Amparo à Pesquisa e ao Desenvolvimento Científico e Tecnológico do Maranhão (FAPEMA),  Conselho Nacional de Desenvolvimento Cient\'ifico e Tecnol\'ogico (CNPq), Coordena\c{c}\~ao de Aperfei\c{c}oamento de Pessoal de N\'ivel Superior (CAPES) -- Finance Code 001, from Brazil, for partial financial support. L.A.L is supported by FAPEMA BPD- 08975/24.
\end{acknowledgments}
\appendix
\section{ Black hole solution after the spontaneous symmetry breaking of Weyl geometry} \label{apen}
In this appendix, we present the complete solution~\eqref{completa} in terms of the coordinate $x$. Substituting Eq.~\eqref{coorde} into Eqs.~\eqref{g00} and~\eqref{g11} yields the following expressions
\begin{widetext}
   \begin{align} \label{A1}
       & \tilde{A}(x) = \frac{9 a_4 M_{Pl}^4 x^4 ( 6a_3+\chi )+ 48 \alpha ^2 M_{Pl}^2 x^2 \left(\left(288 a_2 a_4^2+ \chi  \left(2 (5 a_1 a_4+a_4)-3 a_3^2\right)\right)+12 (1-7 a_1) a_3 a_4+18 a_3^3\right)}{64 \alpha ^4 (- 6a_3 +\chi )^3}
   \end{align}
   \begin{align} \label{B1} \nonumber
     &  \tilde{B}(x) =\frac{(\chi - 6 a_3) \left(4 \alpha ^2 a_3^2+a_4 M_{Pl}^2 x^2\right)}{1728 a_4^3 M_{Pl}^4 x^4}\bigg( \chi  \left(32 \alpha ^2 (5 a_1+1) a_4-48 \alpha ^2 a_3^2+3 a_4 M_{Pl}^2 x^2\right)
     \\& + \left( 96 \alpha ^2 \left(2 (1-7 a_1) a_3 a_4+48 a_2 a_4^2+3 a_3^3\right)+18 a_3 a_4 M_{Pl}^2 x^2 \right) \bigg)
   \end{align}
\end{widetext}
where
\begin{equation}
    \chi = \sqrt{36 a_3^2+3 a_4 \left(32(1-a_1)+\frac{3 M_{Pl}^2 x^2}{\alpha ^2}\right)}. 
\end{equation}Moreover, we use the fact that the $|\frac{dx}{dr}|= \frac{8\alpha}{M_{Pl}}\sqrt{a_4+\frac{4\alpha^2a_3^2}{M_{Pl}^2 x^2}}$.

It is fundamentally important to recover General Relativity within our framework. However, a difficulty arises: this limit would seemingly be achieved when the scalar field is constant (i.e., $a_1 = 1$ and $a_3 = 0$) and when $\alpha = 0$. Yet, in the latter case, we encounter a divergent cosmological constant term. To circumvent this issue, we will impose only the first condition ($a_1 = 1$ and $a_3 = 0$), which corresponds to freezing the scalar field dynamics. Under this condition, we obtain the following solution
\begin{equation} \label{sch}
   \tilde{A}(x)=\tilde{B}(x)^{-1} = 1 +\frac{8  a_2 \sqrt{a_4}\alpha}{M_{Pl} x} +\frac{M_{Pl}^2 x^2}{64 \alpha^2}. 
\end{equation}
Therefore, it is possible to show that in this limit, we can to recover the Schwarzschild-de Sitter solution, which, as is well known, is a vacuum solution of Einstein gravity.

Assuming that $a_2= -R_{Sc}$, $a_3= \frac{\tilde{q}^2}{12R_{Sc}}$ and $a_4=\frac{M_{Pl}^2}{64\alpha^2}$ into Eqs.~\eqref{A1} and~\eqref{B1}, we obtain that
\begin{widetext}
   \begin{align} \label{Amodif}
       & \tilde{A}(x) =  \frac{-32 \tilde{q}^4 x^2+8 \zeta  R_{Sc} \left(\tilde{q}^2 x^2-24 R_{Sc}^2\right)-768 \tilde{q}^2 R_{Sc}^2+9 R_{Sc}^2 x^2 \epsilon ^2 \left(x^2 \epsilon ^2+64\right)}{64 \left(-\frac{4 \tilde{q}^2}{\epsilon ^2}+\zeta  R_{Sc}\right)^2}
   \end{align}
     \begin{align} \label{Bmodif} 
       & \tilde{B}(x) =\left(\frac{\tilde{q}^4}{324 R_{Sc}^4 x^2 \epsilon ^6}+\frac{1}{576 R_{Sc}^2 \epsilon ^2}\right)\bigg( \frac{ 9 \zeta  R_{Sc}^3 \epsilon ^2 \left(x^2 \epsilon ^2+64\right)+(256 \tilde{q}^6-64 \zeta  \tilde{q}^4 R_{Sc}+36 q^2 R_{Sc}^2 \epsilon ^2 \left(x^2 \epsilon ^2-64\right)-1728 R_{Sc}^4 \epsilon ^4)}{-4 \tilde{q}^2+\zeta  R_{Sc}}\bigg)
   \end{align}
\end{widetext}  
where $\epsilon = \frac{M_{Pl}}{\alpha}$ and
\begin{equation}
    \zeta   = \sqrt{\frac{16 \tilde{q}^4}{R_{Sc}^2}+9 x^2 \epsilon ^4}.
\end{equation}

\bibliography{refs.bib}
\bibliographystyle{report}
\end{document}